\documentclass[12pt]{article}

\begin{document}

\title{Gravitation as a Quantum Diffusion\footnote{Published in: ''Z.Zakir (2003) \textit{Structure of Space-Time and Matter}, CTPA, Tashkent.'' }}
\author{Zahid Zakir\thanks{E-mail: zahid@in.edu.uz}\\Centre for Theoretical Physics and Astrophysics,\\ P.O.Box 4412, Tashkent 700000 Uzbekistan}
\date{June 17, 1999;\\
Revised: October 17, 2003.}
\maketitle

\begin{abstract}
Inhomogeneous Nelson's diffusion in flat spacetime with a tensor of diffusion
can be described as a homogeneous one in a Riemannian manifold with this
tensor of diffusion as a metric tensor. The influence of matter to the energy
density of the stochastic background (vacuum) is considered. It is shown that
gravitation can be represented as inhomogeneity of the quantum diffusion, the
Einstein equations for the metrics can be derived as the equations for the
corresponding tensor of diffusion.
\end{abstract}

\section{Introduction}

In \cite{Za1} inhomogeneous conservative diffusion with the tensor of
diffusion $\nu^{ab}(x,t)$ has been considered, and the similarities in the
geometric descriptions of gravitation and quantum fluctuations have been
discussed. The proposed idea about the physical nature of gravitation is
that\textit{\ gravitation is inhomogeneous Nelson's diffusion, i.e. a
consequence of the quantum fluctuations. }This treatment may be applied also
to some gauge fields by means of the Kaluza-Klein mechanism.

In \cite{Za1} only the behavior of a sample particle on inhomogeneous
stochastic background has been described. In this paper the influence of the
source to this stochastic background will be considered, and gravitation will
be described fully in terms of the general quantum diffusion. Then the
Einstein equations will be related with the corresponding diffusion equations.

\section{\textbf{Sample particles on inhomogeneous stochastic background}}

At inhomogeneous diffusion, considered in \cite{Za1}, the mean acceleration
$E[a^{i}(x,t)]$ does not contain terms with the osmotic velocity $u^{i}$, and
for the diffusion of the free particle we have:%

\begin{equation}
E\left[  \frac{\partial v_{i}}{\partial t}+(\mathbf{v\nabla})v_{i}\right]  =0.
\end{equation}
Here a new diffusional acceleration appears due to the presence of derivatives
of the metrics in the Laplace-Beltrami operator $\mathbf{\nabla}$:%

\begin{equation}
E\left[  \frac{\partial v_{i}}{\partial t}+(\mathbf{v\nabla})v_{i}\right]
=E[\Gamma_{ij}^{k}v^{j}v_{k}].
\end{equation}
The diffusional acceleration $E[\Gamma_{ij}^{k}v^{j}v_{k}]$ does not depend on
the mass $m$ of the sample particle, i.e. we have an exact analog of the
equivalence principle.

L.Smolin \cite{Sm} had paid attention to the equality of the quantum
diffusional mass $m_{q}$, determining from the diffusion coefficient
$\nu=\hbar/2m_{q},$ and the inertial mass $m_{in}$ with the high accuracy
$(m_{in}-m_{q})/m_{in}<4\times10^{-13}$. Here we can consider this fact as
following from \textit{the generalized equivalence principle} - the
equivalence between the inertial motion in curved spacetime, the motion in the
gravitational field and inhomogeneous quantum diffusion.

Independence of the acceleration on the mass of the sample particle leads also
to the same acceleration of macroscopic objects and the basises of reference
frames. The acceleration of the reference frame means the appearance of
non-trivial macroscopic metric and non-zero curvature of space-time.

\section{The influence of matter to the stochastic background}

At the interaction with the stochastic background, a massive classical
particle of a bare mass $m_{0}$ undergoes the stochastic fluctuations. At
these fluctuations the energy of the vacuum around the particle partly
transforms into the energy of particle's fluctuations. As the result,
\textit{particle's energy increases} to the quantum fluctuations energy
$T_{ik}^{(q)}(m_{0}^{-1})$ which is inverse proportional to the bare mass
$m_{0}$, and the physical energy momentum density of matter becomes equal to:%

\begin{equation}
T_{ik}(m)=T_{ik}^{(0)}(m_{0})+T_{ik}^{(q)}(\frac{1}{m_{0}}).
\end{equation}

Due to this energy transfer \textit{the vacuum energy density around the
source decreases} with respect to the unperturbed vacuum at the spatial
infinity. This lowering of the vacuum energy density is maximal near the
source and vanish at large distances. The such \textit{inhomogeneity of the
vacuum energy density leads to a coordinate dependence of the diffusion
coefficient} $\nu_{ik}(x)$ which can be considered as the metric tensor of the
effective Riemannian manifold. As it was shown in \cite{Za1}, the lowering of
the energy density near the massive object leads to the diffusional
acceleration of sample particles to that source so that this acceleration does
not depend on the mass of the sample particles.

As a physical model of this effect we can consider the behavior of two
classical particles in an accelerated frame of reference. Let the first
particle's trajectory be a geodesic line. This particle does not interact with
the frame of reference and its ''gravitational energy'' is zero. Let the
second particle be accelerated together with one of the local frames of the
accelerated frame, and it is at rest in this frame. In this case the local
frame expends the energy for the acceleration of the particle. As a result,
the energy of the particle increases while the energy of the local frame
interacting with the particle decreases. If the local frames at some surface
of the extended reference frame are bonded by elastic springs (same as a
trampoline), then the acceleration of the particle with one of the local
frames leads to the formation of a smooth curved surface around this frame.

Thus, a gravitational energy density is related with the lowering of the
vacuum energy density around the source, i.e. by the decreasing of the
intensity of the quantum fluctuations. This fact we can take into account in
the standard action function $A$:%

\begin{equation}
A=\frac{1}{2}\int d\Omega\sqrt{-\gamma}\left(  -\frac{1}{\kappa}%
R+L_{(m)}\right)  ,
\end{equation}
by the variation not the Lagrangian with the Ricci tensor $R_{ik}$, but
directly containing the Riemann tensor $R_{ilkm}$ as in \cite{Za2}. We have:%

\begin{equation}
\delta A=\frac{1}{2}\int d\Omega\sqrt{-\gamma}[G_{iklm}+T_{iklm}]\gamma
^{il}\delta\gamma^{km}=0,
\end{equation}
and%

\begin{equation}
G_{iklm}+T_{iklm}=0.
\end{equation}
Here:%
\begin{equation}
G_{iklm}=\frac{1}{\kappa}[R_{iklm}-\frac{1}{6}(\gamma_{il}\gamma_{km}%
-\gamma_{im}\gamma_{kl})R],
\end{equation}%

\begin{equation}
T_{iklm}=V_{iklm}+\frac{1}{2}(\gamma_{km}T_{il}-\gamma_{kl}T_{im}+\gamma
_{il}T_{km}-\gamma_{im}T_{kl})-\frac{1}{6}(\gamma_{il}\gamma_{km}-\gamma
_{im}\gamma_{kl})T.
\end{equation}

The energy-momentum density tensor of the source $T_{iklm}$ contains the
energy-momentum density of the matter $T_{ik}$, its scalar $T=g^{ik}T_{ik}$,
and the new term $V_{iklm}$ which is the \textit{4-index energy-momentum
density tensor for the gravitational field} \cite{Za2} with zero contraction
$g^{il}V_{iklm}=0$. In the vacuum the $R_{iklm}$ is equal to the Weyl tensor
$C_{iklm}$ which, in fact, determines the energy-momentum density of the
gravitational field as:
\begin{equation}
\frac{1}{\kappa}C_{iklm}=V_{iklm}.
\end{equation}

In the asymptotically flat spacetime this definition of the gravitational
energy leads to the same total energy of the source and its gravitational
field as the pseudotensor and Hamiltonian approaches.

Another evidence of the quantum diffusional nature of gravitation is the
explanation of the time dilation in the gravitational field. Due to the
slowering of the quantum fluctuations around the massive source (the vacuum
around had ''lost'' the energy for the fluctuating the source), the
frequencies of the wave functions of sample particles and energy levels of
atoms, related with the intensity of the quantum fluctuations, become
redshifted, i.e. all quantum processes near the source occur slower than at
the spatial infinity.

\end{document}